\documentclass[aps,prl,showpacs,twocolumn,floats,superscriptaddress]
{revtex4}
\usepackage{graphicx}
\usepackage{bm}

\begin{document}

\title{Spontaneous spin accumulation in singlet-triplet Josephson
junctions}

\author{K. Sengupta}
\affiliation{TCMP division, Saha Institute of Nuclear Physics, 1/AF
Bidhannagar, Kolkata-700064, India}

\author{Victor M. Yakovenko}
\affiliation{Joint Quantum Institute and Center for Nanophysics and
Advanced Materials, Department of Physics, University of Maryland,
College Park, Maryland 20742-4111, USA}

\date{v.10, edited by VMY 12 September 2008, compiled \today}

\begin{abstract}
We study the Andreev bound states in a Josephson junction between a
singlet and a triplet superconductors.  Because of the mismatch in
the spin symmetries of pairing, the energies of the spin up and down
quasiparticles are generally different.  This results in imbalance of
spin populations and net spin accumulation at the junction in
equilibrium.  This effect can be detected using probes of local
magnetic field, such as the scanning SQUID, Hall, and Kerr probes.
It may help to identify potential triplet pairing in $\rm(TMTSF)_2X$,
$\rm Sr_2RuO_4$, and oxypnictides.
\end{abstract}


\maketitle
%%%%%%%%%%%%%%%%%%%%%%%%%%%%%%%%%%%%%%%%%%%%%%%%%%%%%%%%%%%%%%%%%%%%%

Superconductivity with unconventional pairing, particularly
spin-triplet pairing, attracts a lot of interest in the condensed
matter physics community and beyond.  There is significant
experimental evidence in favor of triplet pairing in the
quasi-one-dimensional (Q1D) organic superconductors $\rm(TMTSF)_2X$
\cite{TMTSF,Q1D}, ruthenate $\rm Sr_2RuO_4$ \cite{Maeno01}, and some
heavy-fermion materials.  In the recently discovered oxypnictide
superconductors \cite{FeAs1}, some experiments (observation of
zero-bias conductance peak in tunneling \cite{Shan08} and $H_{c_2}$
exceeding the Pauli paramagnetic limit \cite{Hunte08}) suggest a
possible triplet pairing.  Triplet \cite{Xu08} and singlet
\cite{Mazin08} pairings were proposed in different theoretical models
of oxypnictides.  However, triplet pairing is not firmly established in any these materials.
In this paper, we propose a new physical effect, which can provide useful information about spin
symmetry of superconducting pairing.  We predict that electron spin accumulation should
spontaneously develop at an interface between a singlet and a triplet superconductors.  Other
methods for detection of triplet pairing using superconducting junctions were proposed in
Refs.~\cite{Tanaka07,Asano07}.

The predicted spin accumulation originates from the Andreev bound
states at the interface between the singlet and triplet
superconductors.  Because of the mismatch between the spin symmetries
of pairing, the spin up and down Andreev bound states have different
energies.  This results in different population of the spin up and
down states and net spin accumulation at the interface.  The preferred
axis for the spin projection is determined by the vector $\bm d$ of
the triplet pairing, and the sign of the accumulated spin is
determined by spontaneous symmetry breaking.  The resulting
magnetization can be detected using local probes of magnetic field,
such as the scanning SQUID, Hall, or Kerr probes.  If a voltage $V$ is
applied to the junction, then the magnetization would oscillate with
the Josephson frequency $2eV/\hbar$.  In this paper, we present
calculations in two cases: for the non-chiral $p_x$-wave pairing,
relevant to the $\rm(TMTSF)_2X$ materials \cite{Sengupta01}, and for
the chiral $p_x+ip_y$ pairing, relevant to $\rm Sr_2RuO_4$
\cite{Sengupta02}.  For oxypnictides, the calculations would be more
complicated because of the multiple bands \cite{Mazin08}, but the
result should be qualitatively the same.  Although the Andreev bound
states \cite{Asano03} in the singlet-triplet Josephson junctions
\cite{Yip93} were studied in literature before, the spin accumulation
effect was not recognized, except in Ref.~\cite{Kwon04} for the
special case of equal energy gaps.

\begin{figure}
\includegraphics*[width=0.73\linewidth]{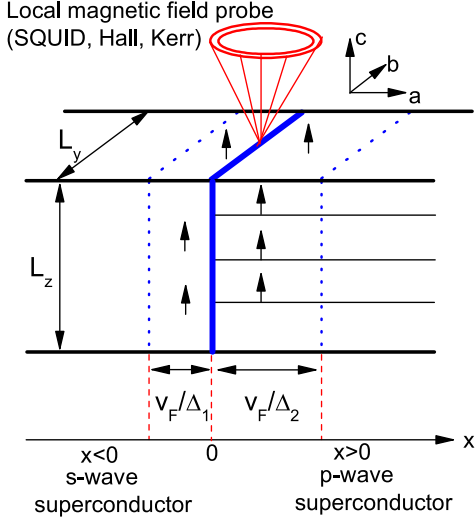}
\caption{A Josephson junction between a single and a triplet
  superconductors.  The thin black solid lines represent chains for
  $\rm(TMTSF)_2X$ or planes for $\rm Sr_2RuO_4$.  The blue dotted
  lines indicates the localization lengths of the Andreev bound
  states.  The cone represents a probe of the local magnetic field
  produced by spin accumulation at the interface.}
\label{fig:setup}
\end{figure}

Let us consider a Josephson junction between an $s$-wave and a
$p$-wave superconductors located at $x<0$ and $x>0$ respectively, as
shown in Fig.~\ref{fig:setup}.  The two superconductors are separated
by a narrow insulating barrier at $x=0$, which is modeled by the
delta-function potential $U(x)= U_0\delta(x)$.  The interface between
the superconductors is assumed to be smooth, so that the electron
momentum parallel to the interface $\bm k_\|=(k_y,k_z)$ is a good
quantum number.  The interface plane is perpendicular to the planes of
$\rm Sr_2RuO_4$ or chains of $\rm(TMTSF)_2X$, as shown in
Fig.~\ref{fig:setup}.  The pairing potential of the singlet $s$-wave
superconductor on the left ($L$) side is
\begin{equation}
  \left<\hat c_{\bm k,\sigma} \hat c_{-\bm k,\bar\sigma}\right>
  \propto \Delta_1 {\rm sgn}(\sigma) \equiv \Delta_\sigma^L,
\label{Delta_s}
\end{equation}
where $\hat c_{\bm k,\sigma}$ is the destruction operator of an
electron with the momentum $\bm k$ and spin
$\sigma=\uparrow,\downarrow$.  Here, ${\rm sgn}(\sigma)=+(-)$ and
$\bar\sigma=\downarrow(\uparrow)$ for $\sigma=\uparrow(\downarrow)$.
The pairing potential of the triplet $p$-wave superconductor on the
right ($R$) side is
\begin{equation}
  \left<\hat c_{\bm k,\sigma} \hat c_{-\bm k,\sigma'}\right>
  \propto
  \Delta_2 \, i\hat\sigma_y (\hat{\bm\sigma}\cdot\bm d) \, f(\bm k) \,
  e^{i\phi_0}.
\label{pp1}
\end{equation}
Here $\hat{\bm\sigma}$ are the Pauli matrices in the spin space, $\bm
d$ is a unit vector characterizing the spin polarization of the
triplet superconductor, and $\phi_0$ is the U(1) phase difference
across the junction.  The function $f(\bm k)$ represents the orbital
symmetry of the pairing potential: $f(\bm k)=(k_x+ik_y)/k_F$ for the
chiral $p_x+ip_y$ pairing and $f(\bm k)=k_x/k_F$ for the non-chiral
$p_x$ pairing, where $k_F$ is the Fermi momentum.  We assume that the
vector $\bm d$ has a uniform orientation independent of $\bm k$.  By
selecting the spin quantization axis $\hat z$ along $\bm d$,
Eq.~(\ref{pp1}) is simplified as
\begin{equation}
  \left<\hat c_{\bm k,\sigma} \hat c_{-\bm k,\bar\sigma}\right>
  \propto \Delta_2 \, f(\bm k) \, e^{i\phi_0}
  \equiv \Delta_\sigma^R.
\label{Delta_p}
\end{equation}
In this representation, both singlet (\ref{Delta_s}) and triplet
(\ref{Delta_p}) pairing potentials couple electrons with opposite
spins.

%%%%%%%%%%%%%%%%%%%%%%%%%%%%%%%%%%%%%%%%%%%%%%%%%%%%%%%%%%%%%%%%
%%%%%%%%%%%%%%%%%%%%%%%%%%%%%%%%%%%%%%%%%%%%%%%%%%%%%%%%%%%%%%%%

Electron states in a superconductor are described by the Bogoliubov
operators $\hat\gamma$, which are related to the electron operators
$\hat c$ by the following equations \cite{Zagoskin}
\begin{eqnarray}
  \!\!\!\!\!\!\!\!\!\!\!\!\!\!
  && \hat\gamma_{\sigma n \bm k_\|} = \int dx\,
  [u_{\sigma n \bm k_\|}^*(x) \, \hat{c}_{\sigma \bm k_\|}(x)
  +v_{\sigma n \bm k_\|}^*(x) \,
  \hat{c}_{\bar\sigma \bar{\bm k}_\|}^\dag(x)],
\label{gamma_n} \\
  \!\!\!\!\!\!\!\!\!\!\!\!\!\!
  && \hat c_{\sigma \bm k_\|}(x) = \sum_{n}
  [u_{\sigma n \bm k_\|}(x) \, \hat\gamma_{\sigma n \bm k_\|}
  + v_{n \bar\sigma \bar{\bm k}_\|}^*(x) \,
  \hat\gamma_{n \bar\sigma \bar{\bm k}_\|}^\dag],
\label{canon}
\end{eqnarray}
where $\bar{\bm k}_\|=-\bm k_\|$, and $n$ is the quantum number of the
Bogoliubov eigenstates.  The two-component wave functions
$\psi_{\sigma n}(x,\bm k_\|)=[u_{\sigma n \bm k_\|}(x),v_{\sigma n \bm
k_\|}(x)]$ are the eigenstates of the Bogoliubov-de Gennes (BdG)
equation with the eigenenergies $E_{\sigma n \bm k_\|}$
\begin{equation}
  \left(\begin{array}{cc}
  \hat H_0 + U(x) & \Delta_\sigma(x,\hat k_x,\bm k_\|) \\
  \Delta_\sigma^*(x,\hat k_x,\bm k_\|) & -\hat H_0-U(x)
  \end{array}\right) \psi_{\sigma}(x)
  = E_{\sigma} \psi_{\sigma}(x),
\label{bdg1}
\end{equation}
where we omitted the labels $n$ and $\bm k_\|$ for $\psi$ and $E$ to
shorten notation.  Here $\hat H_0=\varepsilon(\hat k_x,\bm k_\|)-\mu$
is the electron dispersion relation with $\hat k_x=-i\partial_x$ and
the chemical potential $\mu$, and we set $\hbar=1$.  In $\rm
Sr_2RuO_4$, the main Fermi surface is circular, so we take
$\varepsilon=(k_x^2+k_y^2)/2m$, where $m$ is the effective mass.  In
Q1D conductors, the Fermi surface consists of two open sheets
perpendicular to the chains, so we take
$\varepsilon=k_x^2/2m-t_b\cos(bk_y)-t_c\cos(ck_z)$, where the first
term represents motion along the chains, and $t_b$, $t_c$, $b$, and
$c$ are the interchain tunneling amplitudes and spacings \cite{t_c}.
Notice that the spin projection $\sigma$ is a good quantum number for
the Bogoliubov quasiparticles (\ref{gamma_n}), and the BdG equations
(\ref{bdg1}) separate for $\sigma=\uparrow$ and $\downarrow$.  The
pairing potential $\Delta_\sigma$ in Eq.~(\ref{bdg1}) is given by
Eq.~(\ref{Delta_s}) for $x<0$ and by Eq.~(\ref{Delta_p}) for $x>0$.

The wave functions $\psi_\sigma^L$ and $\psi_\sigma^R$ on the left and
right sides of the junction satisfy the standard boundary condition at
$x=0$ obtained by integrating Eq.~(\ref{bdg1}) over $x$ from $-0$ to
$+0$
\begin{equation}
  \psi_\sigma^R(0)=\psi_\sigma^L(0), \;
  (\partial_x\psi_\sigma^R - \partial_x\psi_\sigma^L)_{x=0}
  = 2mU_0 \psi_\sigma^L(0).
\label{x=0}
\end{equation}
We are interested in the subgap bound states of Eq.~(\ref{bdg1}) with
energies $|E|\le{\rm Min}[\Delta_1,\Delta_2]$, which are localized
near the junction.  Such localized solutions can be obtained as a
superposition of the wavefunctions for the right and left moving
quasiparticles \cite{Kwon04}:
\begin{equation}
  \psi_\sigma^\beta(x) =
  e^{-\kappa_\sigma^\beta x}
  \Bigg[ A^\beta e^{i\tilde k_Fx} \left(\begin{array}{c}
  u_{\sigma+}^\beta \\ v_{\sigma+}^\beta
  \end{array}\right)
 + B^\beta e^{-i\tilde k_F x} \left(\begin{array}{c}
  u_{\sigma-}^\beta \\ v_{\sigma-}^\beta
  \end{array}\right) \Bigg].
\label{wavefn1}
\end{equation}
Here the superscript $\beta=R,L$ labels the wave functions on the
right and left sides of the junction.  The subscript $\alpha=\pm$ in
$u_{\sigma\alpha}$ or $v_{\sigma\alpha}$ denotes the right
($\alpha=+$) or left ($\alpha=-$) moving quasiparticles.  The
parameters $\kappa_\sigma^{R(L)}=+(-)\sqrt{\Delta_{2(1)}^2
-E_\sigma^2}/\tilde v_F$ determine the inverse localization lengths of
the bound states inside the right and left superconductors.  The
variables $\tilde k_F$ and $\tilde v_F$ are the $x$ components of the
Fermi momentum and Fermi velocity, which, generally, depend on $\bm
k_\|$.  For $\rm Sr_2RuO_4$, $\tilde k_F=\sqrt{k_F^2-k_y^2}$ and
$\tilde v_F=\tilde k_F/m$.  For Q1D conductors, $\tilde
k_F=k_F+2t_b\cos(bk_y)/v_F+2t_c\cos(ck_z)/v_F$ and $\tilde v_F\approx
v_F$, where $v_F=k_F/m$.  The coefficients $u_{\sigma\alpha}^\beta$
and $v_{\sigma\alpha}^\beta$ are determined by substituting the
right and left moving terms into Eq.\ (\ref{bdg1}) away from the
junction.  They satisfy
\begin{equation}
  \eta_{\sigma\alpha}^\beta =
  \frac{u_{\sigma\alpha}^\beta}{v_{\sigma\alpha}^\beta}
  = \frac{E_\sigma -
  i\,{\rm sgn}(\alpha)\,\tilde v_F\kappa_\sigma^\beta}
  {\Delta_\sigma^\beta},
\label{frac1}
\end{equation}
where $\Delta_\sigma^\beta$ are given by Eqs.~(\ref{Delta_s}) and
(\ref{Delta_p}).  Notice that all variables in Eqs.~(\ref{wavefn1})
and (\ref{frac1}), generally, depend on $\bm k_\|$.

Substituting Eq.~(\ref{wavefn1}) into Eq.~(\ref{x=0}), we obtain a set
of 4 linear homogeneous equations for the coefficients $A^\beta$ and
$B^\beta$.  A condition for non-zero solutions requires vanishing of
the determinant of the corresponding $4\times4$ matrix, which yields
the following equation
\begin{equation}
  \frac{(\eta^R_{\sigma+} - \eta^L_{\sigma+})
  (\eta^R_{\sigma-} - \eta^L_{\sigma-})}
  {(\eta^R_{\sigma-} - \eta^L_{\sigma+})
  (\eta^R_{\sigma+} - \eta^L_{\sigma-})} = 1-D(\bm k_\|).
\label{cond1}
\end{equation}
Here $D(\bm k_\|)=4/(4+Z^2(\bm k_\|)$ is the transmission coefficient
of the barrier.  For Q1D conductors, $Z(\bm k_\|)=Z_0=2U_0/v_F$ is
independent of the transverse momentum.  For $\rm Sr_2RuO_4$, $Z(\bm
k_\|)=Z_0k_F/\sqrt{k_F^2-k_y^2}$.  Substituting Eq.~(\ref{frac1}) into
Eq.~(\ref{cond1}), we obtain an equation for the energies of the
Andreev bound states
\begin{equation}
  {\mathcal A} \sin(\Phi_{\bm k_\|}) - {\mathcal B}\cos(\Phi_{\bm k_
\|})
  = {\rm sgn}(\sigma) \, \sin(\phi_0) \, D(\bm k_\|).
\label{andreq}
\end{equation}
Here $\phi(\bm k)$ is the phase of the function $f(\bm k)$ in
Eq.~(\ref{Delta_p}), and $\Phi_{\bm k_\|}=[\phi(\tilde k_F,\bm
k_\|)-\phi(-\tilde k_F,\bm k_\|)]/2$ is a half of the phase
difference
between the points on the Fermi surface connected by specular
reflection from the barrier, selected so that $0\leq\Phi\leq\pi$
\cite{Sengupta02}.  The coefficients $\mathcal A$ and $\mathcal B$ are
\begin{eqnarray}
  {\mathcal A} &=& [2-D(\bm k_\|)] \, \epsilon_{\sigma1} \,
  \sqrt{1-\epsilon_{\sigma2}^2} + D(\bm k_\|) \,
  \epsilon_{\sigma2} \, \sqrt{1-\epsilon_{\sigma1}^2},
\nonumber\\
  {\mathcal B} &=& -D(\bm k_\|) \, \epsilon_{\sigma1} \,
  \epsilon_{\sigma2} + [2-D(\bm k_\|)] \,
  \sqrt{(1-\epsilon_{\sigma1}^2)(1-\epsilon_{\sigma2}^2)}.
\nonumber
\end{eqnarray}
where $\epsilon_{\sigma1(2)}=E_\sigma(\bm k_\|;\phi_0)/\Delta_{1(2)}
\le 1$ is the dimensionless energy of the bound state.  In the case
$\Delta_1=\Delta_2$, the equations simplify and reproduce the results
of Ref.\ \cite{Kwon04}.

Notice that $E_\sigma$ depends on the spin index $\sigma$ only through
the right-hand side of Eq.~(\ref{andreq}), which is invariant under the transformation $\sigma\to\bar\sigma$ and $\phi_0\to\phi_0+\pi$, so
$E_\sigma(\bm k_\|;\phi)=E_{\bar\sigma}(\bm k_\|;\phi+\pi)$.  This
relation can be understood by noting that the singlet pairing
potential (\ref{Delta_s}) has opposite signs for $\sigma=\uparrow$ and
$\downarrow$, whereas the triplet pairing potential (\ref{Delta_p})
has the same sign.  So, a phase difference $\phi_0$ across the
junction for spin-up quasiparticles implies the effective phase
difference $\phi_0+\pi$ for spin-down quasiparticles, which explains
the above-mentioned invariance.  We see that the bound states
energies $E_\uparrow$ and $E_\downarrow$ are generally different for a given $\phi_0$, which is the key point for understanding of spin
accumulation \cite{spin-current}.

%%%%%%%%%%%%%%%%%%%%%%%%%%%%%%%%%%%%%%%%%%%%%%%%%%%%%%%%%%%%%%%%
\begin{figure}
\includegraphics*[width=0.8\linewidth]{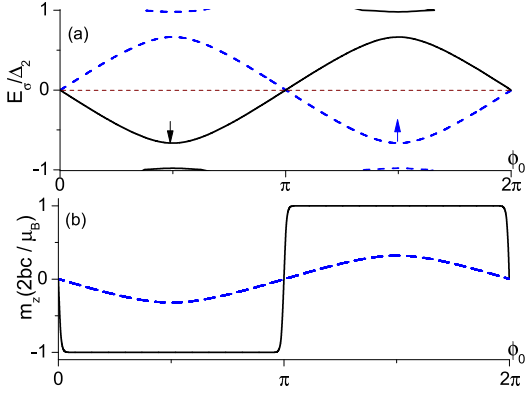}
\caption{(a) The energies $E_\uparrow$ (dashed lines) and
  $E_\downarrow$ (solid lines) of the Andreev bound states vs.\ the
  phase $\phi_0$ between the $s$ and $p_x$-wave superconductors
  ($D=0.8$ and $\Delta_2=0.5\Delta_1$).  (b) Magnetic moment $m_z$
per
  unit area of the interface, Eq.~(\ref{m_z}), vs.\ $\phi_0$ for
  $T/\Delta_2=0.01$ (solid line) and $1$ (dashed line).}
\label{fig:p_x}
\end{figure}
%%%%%%%%%%%%%%%%%%%%%%%%%%%%%%%%%%%%%%%%%%%%%%%%%%%%%%%%%%%%%%%%%

For the $p_x$-wave pairing, we have $f(-k_F)=-f(k_F)$, so $\Phi_{\bm
k_\|}=\pi/2$, and Eq.~(\ref{andreq}) reduces to ${\mathcal A}=D{\rm
sgn}(\sigma)\sin(\phi_0)$, where $D=4/(4+Z_0^2)$.  In this case, the
energies $E_\sigma$ are independent of $\bm k_\|$ and are plotted vs.\
$\phi_0$ in Fig.~\ref{fig:p_x}a.  Depending on $\phi_0$, there are two
or four of such states for each $\bm k_\|$.  The spin up and down
bound states have opposite energies $E_\uparrow=-E_\downarrow$, as
shown by the dashed and solid lines in Fig.~\ref{fig:p_x}a.  The
difference between the Fermi populations of the spin up and down
states gives a net magnetic moment $m_z$ per unit area of the
interface \cite{center}
\begin{equation}
  m_z = \frac{\mu_B}{2bc} \,
  \tanh\left(\frac{E_{\downarrow}(\phi_0)}{2T}\right),
\label{m_z}
\end{equation}
where $\mu_B$ is the Bohr magneton, $T$ is the temperature, and the
prefactor 1/2 compensates for double-counting \cite{Sengupta01,Kwon04}.  A plot of $m_z$ vs.\ $\phi_0$ is shown in Fig.~\ref{fig:p_x}b for two different temperatures.  At low $T$, the magnetic moment is close to $\mu_B/2$ per chain, since only the lower energy state is populated.  In an open circuit, the value of the phase $\phi_0$ is determined by minimization of the total energy of the system.  For the energy levels shown in Fig.~\ref{fig:p_x}a, the minimum is achieved at either $\phi_0=\pi/2$ or $\phi_0=3\pi/2$ (the same as $\phi_0=-\pi/2$) \cite{odd-freq}.  The system spontaneously breaks the symmetry and selects one of the two energy minima with negative or positive magnetization.

%%%%%%%%%%%%%%%%%%%%%%%%%%%%%%%%%%%%%%%%%%%%%%%%%%%%%%%%%%%%%%%%
\begin{figure}
\includegraphics*[width=0.8\linewidth]{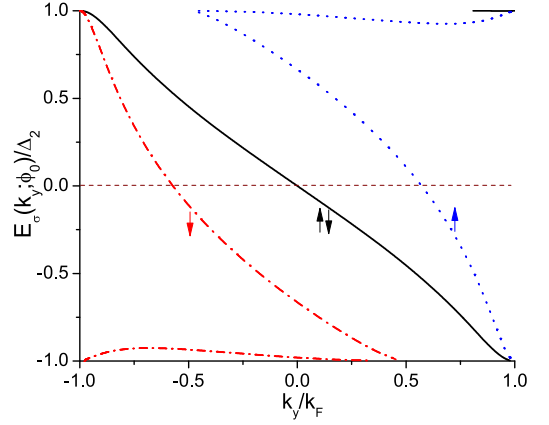}
\caption{The energies $E_\uparrow(k_y)$ (blue dotted lines) and
  $E_\downarrow(k_y)$ (red dash-dotted lines) vs.\ $k_y$ for the phase
  $\phi_0=\pi/2$ between the $s$ and $p_x+ip_y$ superconductors
  ($Z_0=1$).  For $\phi_0=0$ and $\pi$,
  $E_\uparrow(k_y)=E_\downarrow(k_y)$ is shown by the black solid
  line.}
\label{fig:E}
\end{figure}
%%%%%%%%%%%%%%%%%%%%%%%%%%%%%%%%%%%%%%%%%%%%%%%%%%%%%%%%%%%%%%%%

\begin{figure}
\includegraphics*[width=0.8\linewidth]{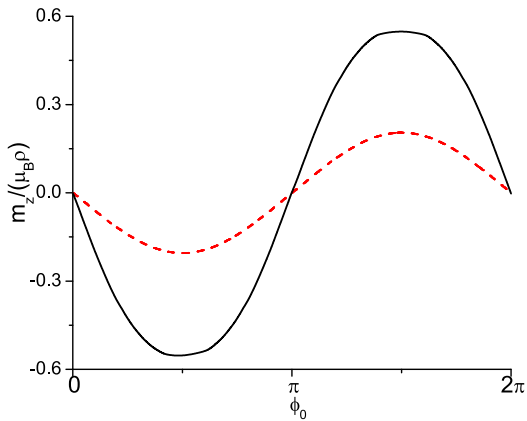}
\caption{Normalized magnetic moment $m_z$ per unit area of the
  interface, Eq.~(\ref{magchiral}), vs.\ $\phi_0$ for $T=0.01\Delta_2$
  (solid line) and $T=\Delta_2$ (dashed line) for a junction between
  the $s$ and $p_x+ip_y$ superconductors ($\rho=k_F/2\pi c$).}
\label{fig:m_z}
\end{figure}

For the chiral $p_x+ip_y$ pairing, we have $\phi(\bm
k)=\arctan(k_y/k_x)$ and $\Phi_{k_\|}=\phi(k_y)+\pi/2$, so that
$\sin\Phi_{k_\|}=|k_x|/k_F$ and $\cos\Phi_{k_\|}=-k_y/k_F$.  Then,
Eq.~(\ref{andreq}) gives the energies $E_\sigma(k_y;\phi_0)$ dependent on the transverse momentum $k_y$ because of the broken time-reversal
symmetry.  The plot of $E_\sigma(k_y;\phi_0)$ vs.\ $k_y$ for several
values of $\phi_0$ is shown in Fig.\ \ref{fig:E}.  The energy
splitting between $E_\uparrow$ and $E_\downarrow$ is maximal at
$\phi_0=\pm\pi/2$ and vanishes at $\phi_0=0$ and $\pi$.  The
imbalance between the spin up and down populations produces the net magnetic moment $m_z$ per unit area of the interface
\begin{equation}
  m_z = \frac{\mu_B}{2c} \sum_{\sigma=\uparrow,\downarrow}
  {\rm sgn}(\sigma)
  \int\limits_{-k_F}^{k_F} \frac{dk_y}{2\pi} \,
  n_F\left(\frac{E_\sigma(k_y;\phi_0)}{T}\right),
\label{magchiral}
\end{equation}
where $n_F$ is the Fermi distribution function, and $c$ is the
interplane distance for $\rm Sr_2RuO_4$ \cite{center}.  The plot of
$m_z$ vs.\ $\phi_0$ is shown in Fig.\ \ref{fig:m_z} for two different
temperatures.  The minimum of energy is achieved at $\phi_0=\pm\pi/2$,
so the system spontaneous breaks the symmetry and selects one of the
two optimal values for $\phi_0$.

Figs.~\ref{fig:p_x}b and \ref{fig:m_z} show that the magnetic moment
$m_z$ changes sign when $\phi_0$ crosses $\pi$.  If a bias voltage $V$
is applied across the junction, it would make the phase difference
time-dependent: $\phi_0(t)=2eVt$.  Then, the magnetization $m_z$ at
the interface would oscillate with the time period $\pi/eV$. We assume
that the oscillations are slow enough for the spin population to
remain close to the thermal equilibrium at each moment of time.

%%%%%%%%%%%%%%%%%%%%%%%%%%%%%%%%%%%%%%%%%%%%%%%%%%%%%%%%%%%%%%%%
%%%%%%%%%%%%%%%%%%%%%%%%%%%%%%%%%%%%%%%%%%%%%%%%%%%%%%%%%%%%%%%%

Finally, we discuss possible experiments for detection of the spontaneous spin polarization.  A schematic experimental setup is shown in
Fig.~\ref{fig:setup}.  Assuming semi-infinite geometry in the $z$
direction, the magnetization at the junction and the magnetic field
can be estimated as $B\simeq\mu_0\mu_B\rho\kappa/4$, where
$\kappa\simeq\Delta_2/v_F$ is the inverse penetration depth of the
bound states in the case $\Delta_1\gg\Delta_2$, $\rho=1/bc$ for
$\rm(TMTSF)_2X$, and $\rho=k_F/2\pi c$ for $\rm Sr_2RuO_4$.  This
magnetic field can be measured using a scanning SQUID or Hall
microscope \cite{Moller05,Moller07}.  For $\rm(TMTSF)_2X$, we have
$b=0.77$~nm, $c=1.35$~nm and $\kappa^{-1}\simeq0.6$~$\mu$m, which
gives $B\simeq0.3$~G and a magnetic flux $0.06\Phi_0$ (where
$\Phi_0=hc/2e$) through a square scanning SQUID loop of the size
$l=10$~$\mu$m.  For $\rm Sr_2RuO_4$, we have $k_F=7.5\times10^{-9}$~$m^{-1}$, $\kappa^{-1}=66$~nm and $c=1.3$~nm, which gives the field $0.7$~G and the flux $0.15\Phi_0$.  The estimated magnetic fields are well above the typical Hall-probe sensitivity of $80$ mG at $1$ Hz \cite{Moller05}.  However, chiral superconductors are also expected to
have an additional magnetic field due to the charge currents carried
by the chiral Andreev bound states \cite{Kwon03}. A Josephson
junction between $\rm Au_{0.5}In_{0.5}$ and $\rm Sr_2RuO_4$ was scanned using the SQUID and Hall probes in Ref.~\cite{Moller07}, but no spontaneous magnetic field was detected.  A reason for the negative experimental result remains an open question.  Local magnetization can be also detected optically using the Kerr angle rotation \cite{Sih05}.  This effect was observed in Ref.~\cite{Xia06} in the bulk of $\rm Sr_2RuO_4$ due to the orbital time-reversal symmetry breaking in the $p_x+ip_y$ state.  However, the Kerr experiment has not been performed in a scanning mode at a junction with a singlet superconductor.

In conclusion, we have shown that a Josephson junction between a
singlet and a triplet superconductors should exhibit spontaneous spin
accumulation due to mismatch of the spin pairing symmetries.  The
vector of the accumulated spin points along the vector $\bm d$ of the
triplet superconductor.  The sign and magnitude of the spin depend on
the phase difference $\phi_0$ between the superconductors.  In
equilibrium, the system spontaneously breaks symmetry and selects one
of the two values of $\phi_0$ that minimize total energy and maximize
spin accumulation.  When a bias voltage $V$ is applied to the
junctions, the accumulated spin oscillates in time.  The magnetic
field produced by the accumulated spin can be detected using the
SQUID, Hall, or Kerr local probes.

%%%%%%%%%%%%%%%%%%%%%%%%%%%%%%%%%%%%%%%%%%%%%%%%%%%%%%%%%%%%%%%
\vspace{-\baselineskip}

%%%%%%%%%%%%%%%%%%%%%%%%%%%%%%%%%%%%%%%%%%%%%%%%%%%%%%%%%%%%%%%

\begin{thebibliography}{99}
\vspace{-\baselineskip}
%%%%%%%%%%%%%%%%%%%%%%%%%%%%%%%%%%%%%%%%%%%%%%%%%%%%%%%%%%%%%%%

\bibitem{TMTSF} TMTSF stands for tetramethyltetraselenafulvalene and X
represents inorganic anions such ${\rm PF}_6$ or ${\rm ClO}_4$.

\bibitem{Q1D} I.J.~Lee {\it et al.}, Phys. Rev. Lett. {\bf 78}, 3555 (1997);
{\it ibid.} {\bf 88}, 017004 (2001); Phys. Rev. B {\bf 62},   R14669 (2000).

\bibitem{Maeno01} Y. Maeno, T.M. Rice, and M. Sigrist, Phys.  Today
  {\bf 54} (1), 42 (2001); {\bf 54} (3), 104 (2001); A.P.~Mackenzie
  and Y.~Maeno, Rev. Mod. Phys. {\bf 75}, 657 (2003).

\bibitem{FeAs1} Y. Kamihara {\it et al.}, J. Am. Chem. Soc. {\bf 130},
3296 (2008); G.F. Chen {\it et al.}, Phys. Rev. Lett. {\bf 101}, 057007 (2008).

\bibitem{Shan08} L. Shan {\it et al.}, Europhys. Lett. {\bf 83},   57004 (2008).

\bibitem{Hunte08} F. Hunte {\it et al.}, Nature {\bf 453}, 903
(2008).

\bibitem{Xu08} G. Xu {\it et al.}, Europhys. Lett. {\bf 82}, 67002
  (2008); P.A. Lee and X.-G. Wen, arXiv:0804.1739.

\bibitem{Mazin08} I.I. Mazin {\it et al.}, Phys. Rev. Lett. {\bf 101}, 057003 (2008);
K.~Kuroki {\it et al.}, Phys. Rev. Lett. {\bf 101}, 087004 (2008).

\bibitem{Tanaka07} Y. Tanaka {\it et al.}, \prl {\bf 99}, 037005 (2007).

\bibitem{Asano07} Y. Asano {\it et al.}, \prl {\bf 99}, 067005 (2007).

\bibitem{Sengupta01} K. Sengupta {\it et al.}, Phys. Rev. B {\bf
  63}, 144531 (2001).

\bibitem{Sengupta02} K. Sengupta, H.-J. Kwon, and V.M. Yakovenko,
  Phys. Rev. B {\bf 65}, 104504 (2002).

\bibitem{Asano03} Y. Asano {\it et al.}, Phys. Rev. B {\bf 67}, 184505 (2003).

\bibitem{Yip93} S. Yip, J. Low Temp. Phys. {\bf 91}, 203 (1993);
  N. Yoshida {\it et al.}, {\it ibid.} {\bf 117}, 563 (1999);
  Y. Asano {\it et al.}, Phys. Rev. B {\bf 71}, 214501 (2005).

\bibitem{Kwon04} H.-J. Kwon, K. Sengupta, and V.M. Yakovenko,
  Eur. Phys. J. B {\bf 37}, 349 (2004).

\bibitem{Zagoskin} A.M.~Zagoskin, {\it Quantum Theory of Many-Body
  Systems} (Springer, New York, 1998).

\bibitem{t_c} We omit the interplane tunneling term $t_c\cos(ck_z)$
for $\rm Sr_2RuO_4$, because it drops out like for Q1D conductors.

\bibitem{spin-current} The Andreev bound states (\ref{wavefn1}) carry
zero spin current, so there is no spin supercurrent through the
junction.

\bibitem{odd-freq} Somewhat similar behavior was also found for a junction between even-
and odd-frequency  superconductors \cite{Tanaka07}.

\bibitem{center} Eqs.~(\ref{m_z}) and (\ref{magchiral}) omit contributions from states
at the edge of the gap, which have large localization lengths.

\bibitem{Moller05} P.G.~Bjornsson {\it et al.}, Phys. Rev. B {\bf
  72}, 012504 (2005).

\bibitem{Moller07} J.R.~Kirtley {\it et al.}, Phys. Rev. B {\bf 76},
  014526 (2007).

\bibitem{Kwon03} H-J. Kwon, K. Sengupta, and V. M. Yakovenko, Synth.
  Metals {\bf 133-134}, 27 (2003).

\bibitem{Sih05} V. Sih {\it et al.}, Nature Phys. {\bf 1}, 31 (2005).

\bibitem{Xia06} J.~Xia {\it et al.}, Phys. Rev. Lett. {\bf 97},
167002 (2006).

\end{thebibliography}
\end{document}